\documentclass[aps,prb,twocolumn,showpacs,amsmath,amssymb,superscriptaddress]{revtex4-2}
\usepackage{graphicx}
\usepackage{lineno}
\usepackage{CJK}
\usepackage{color}
\usepackage[colorlinks,bookmarks=false,citecolor=blue,linkcolor=red,urlcolor=blue]{hyperref}
\usepackage{multirow}
\usepackage{ulem}
\usepackage{tabularx}
\usepackage[nounderscore]{syntax}
\graphicspath{{figs/}}

\newcommand{\be}{\begin{equation}}
\newcommand{\ee}{\end{equation}}

\begin{document}
\title{Detecting underlying symmetry-protected topological phases via strange correlators and edge engineering
}
\author{Zhe Wang}
\thanks{These authors contributed equally to this work.}
\affiliation{School of Physics and Astronomy, Beijing Normal University, Beijing 100875, China}
\affiliation{Department of Physics, School of Science and Research Center for Industries of the Future, Westlake University, Hangzhou 310030,  China}
\affiliation{Institute of Natural Sciences, Westlake Institute for Advanced Study, Hangzhou 310024, China}
\author{Longye Lu}
\thanks{These authors contributed equally to this work.}
\affiliation{School of Physics and Astronomy, Beijing Normal University, Beijing 100875, China}
\author{Shang-Qiang Ning}
\thanks{These authors contributed equally to this work.}
\affiliation{Department of Physics, The Hong Kong University of Science and Technology, Clear Water Bay, Kowloon, Hong Kong, China} 
\author{Zenan Liu}
\affiliation{Department of Physics, School of Science and Research Center for Industries of the Future, Westlake University, Hangzhou 310030,  China}
\affiliation{Institute of Natural Sciences, Westlake Institute for Advanced Study, Hangzhou 310024, China}

\author{Yan-Cheng Wang}
\email{ycwangphys@buaa.edu.cn}
\affiliation{Hangzhou International Innovation Institute, Beihang University, Hangzhou 311115, China}
\affiliation{Tianmushan Laboratory, Hangzhou 311115, China}

\author{Zheng Yan}
\email{zhengyan@westlake.edu.cn}
\affiliation{Department of Physics, School of Science and Research Center for Industries of the Future, Westlake University, Hangzhou 310030,  China}
\affiliation{Institute of Natural Sciences, Westlake Institute for Advanced Study, Hangzhou 310024, China}

\author{Wenan Guo}
\email{waguo@bnu.edu.cn}
\affiliation{School of Physics and Astronomy, Beijing Normal University, Beijing 100875, China}
\affiliation{Key Laboratory of Multiscale Spin Physics (Ministry of Education), Beijing Normal University, Beijing 100875, China }

\date{\today}
\begin{abstract}
The vast majority of symmetry-protected topological (SPT) states are difficult to detect, which often leads to their misidentification as ordinary or topologically trivial phases. 
In this work, we propose a general framework for detecting these hidden topological states. We distinguish the ordinary matter state from the topological phase by exploiting the boundary 
effects in space (via surface behaviors on engineered edge) and time (via strange correlators) according to the principle of bulk-edge correspondence.
As a concrete example, we study the dimerized spin-1/2 Heisenberg model on a square lattice using quantum Monte Carlo simulations, focusing on its paramagnetic dimer phase and edge states. The dimer phase has been 
widely regarded as topologically trivial due to its gapped edge state on conventional edges. However, the model can also be viewed as 
two-dimensional antiferromagnetically (AF) coupled usual ladders, which suggests an SPT state adiabatically connected to the one-dimensional Haldane phase.
We resolve this puzzle and demonstrate that the dimer phase is indeed a quasi-one-dimensional SPT state  
by measuring generalized strange correlators introduced in this work and by showing that the nontrivial gapless edge state on a zigzag edge is ferromagnetically ordered, resulting from effective ferromagnetic interactions between degenerate spinons liberated on each side of the cut.
Furthermore, we show that the ordered edge state gives rise to an extraordinary surface critical behavior at the (2+1)-dimensional O(3) bulk critical points of the model, which contradicts theoretical predictions based on classical-quantum mapping.
Overall, we establish a standard detection method for uncovering topological phases that masquerade as ordinary states of matter.
\end{abstract}
\maketitle

\section{Introduction}
	\label{intro}
Topological phases of quantum matter are fascinating emergent phenomena commonly characterized by nonlocal order parameters in bulk and exotic behavior at the boundary. A subclass of gapped topological phases is the so-called symmetry-protected topological (SPT) phases, which cannot be smoothly mapped to a product state if only symmetric perturbations are allowed \cite{Chen10, Chen_science, Pollmann2010, 1DSPT, Chen2013CGLWbSPT}.  The most striking feature of the SPT phase is that bulk-gapped quantum phases have gapless or degenerate boundary states as long as the symmetries are not broken, known as the bulk-edge correspondence \cite{physics_of_bTI_2013_ashvin_senthil,bti3,chen14, PhysRevX.6.021015, sto3, sto5, fidkowski13, sto1, sto2, interactingTI_TSc_chong_15, metlitski14,sto4, Fidkowski2018, Lu2012CSSPT,liu2022bulk, PhysRevB.104.075151, PhysRevB.104.075111}.

A simple system with SPT order is the spin-1 chain with antiferromagnetic (AF) Heisenberg interaction (often called the Haldane chain), which has a nonzero excitation gap--the Haldane gap, nonlocal string order, and degenerate edge states~ \cite{Haldane, den1989, Affleck1987, Kennedy1992, Oshikawa_1992, White1996}. Due to the Haldane gap, the topological nature of the Haldane phase in the AF Heisenberg spin-1 chain may remain in (finite) odd-number coupled chains \cite{Pollmann2012} or in quasi-one-dimensional (Q1D) system which is formed by infinite spin-1 chains with weak but finite inter-chain coupling under certain conditions \cite{Matsumoto2001, Moukouri2011, Keola2012, Wierschem2014, Zhu2021}, 
with gapless edge states along the edge perpendicular to the chain direction \cite{Zhu2021}. Experimentally, materials described by Q1D spin-1 or spin-1/2 chains have been found \cite{Ruegg2008,Buyers1986, Hase1993,Honda1998,Uchiyama1999,Hagiwara1998,Hagiwara1998,Okada2016Quasi,Jaime2004Magnetic,Mazurenko2014Nonfrustrated}.

\begin{figure}[htb]
	\centering
	\includegraphics[width=0.5\textwidth]{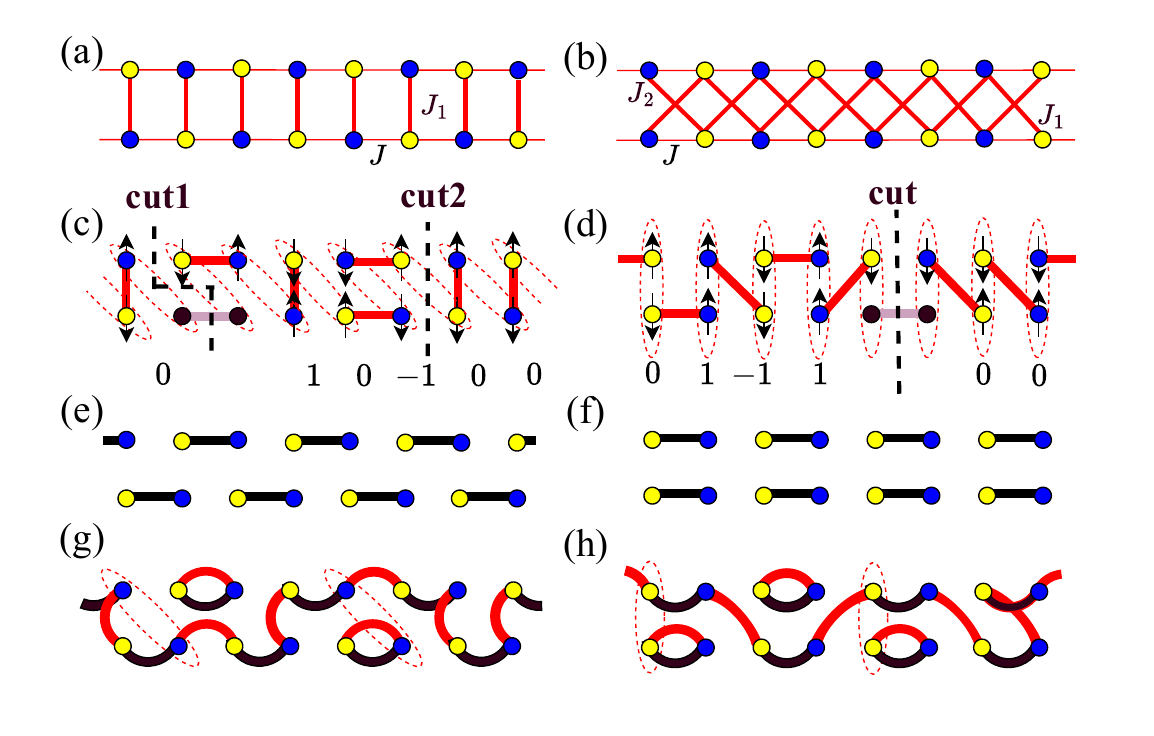}
	\caption{ (a) A usual ladder with AF rung couplings $J_1$ and (b) a diagonal ladder with AF diagonal couplings $J_1$ and $J_2$, typically, $J_1=J_2$.
	$J=1$ denotes the intrachain coupling.
	(c) A particular spin configuration matches a typical short-range valence bond (VB) state, which constitutes the ground state of the usual ladder. 
	Dashed ovals indecate spin-1 $\mathbf{S}^{(\rm E)}_i=\mathbf{S}_{i,0}+\mathbf{S}_{i+1,1}$ 
	After all $S_i^z=0$ removed, the remaining spins show N\'eel order. The thick dashed lines show the two ways to cut systems. Two spinons appear 
	if the cut (cut1) goes through one VB connecting two spin-1s, while cutting the ladder vertically (cut2) breaks a spin-1, no spinon appear. 
	(d) A  particular spin configuration matches a typical short-range VB state that constitutes the ground state of the diagonal ladder. 
	Dashed ovals indicate spin-1 $\mathbf{S}^{(\rm O)}_i=\mathbf{S}_{i,0}+\mathbf{S}_{i,1}$ 
	After all $S_i^z=0$ removed, the remaining spins show N\'eel order.
	The thick dashed line represents a vertical cut that breaks a VB connecting two spin-1s; two spinons are generated.
	The staggered pattern (e) and the columnar pattern (f) of VBs are product states $|\Omega\rangle$ used to define strange correlators. 
	Thick-black lines denote VBs. 
When computing a two-dimensional system, the patterns are repeated along the $y$ direction, covering the lattice model. 
	(g) is the transposition graph of (c) and (e), and (h) is the transposition graph of (d) and (f). 
	}
	\label{Fig:twoladders}
    \end{figure}

	In many cases, the gapped phase of a spin-1/2 two-leg ladder \cite{ElbioDagotto1999} is directly related to the spin-1 Haldane chain with non-zero string order \cite{Kim2000, White1996}. 
 Remarkably, a usual ladder with AF rungs ( as shown in Fig. \ref{Fig:twoladders} (a)) exhibits the Haldane phase similarly 
to that of the diagonal ladder, which has AF interchain couplings on plaquette diagonals ( see Fig. \ref{Fig:twoladders} (b)),  but in different topological sectors.\cite{Kim2000}

Inspired by the Q1D weakly coupled spin-1 chains hosting nontrivial SPT order,  naturally, one expects that the Q1D weakly coupled 
spin-1/2 two-leg ladders also present nontrivial SPT states. Despite the similarity to the  spin-1-chain Q1D 
system, the Q1D spin-1/2 ladders hosts a more sophisticated structure in the UV limit, which opens up more possibilities for studying the 
boundary phenomena that are crucially important for SPT phases.  In the previous works \cite{Wang2022, Wangz2023},  we studied the 
AF and FM coupled diagonal ladders (CDLs) model, which hosts different SPT phases. The SPT character of the two phases was inferred from their adiabatic connection to the 1D Haldane state 
and the presence of gapless edge modes.
    
\begin{figure}[htb]
	\centering
	\includegraphics[width=0.46\textwidth]{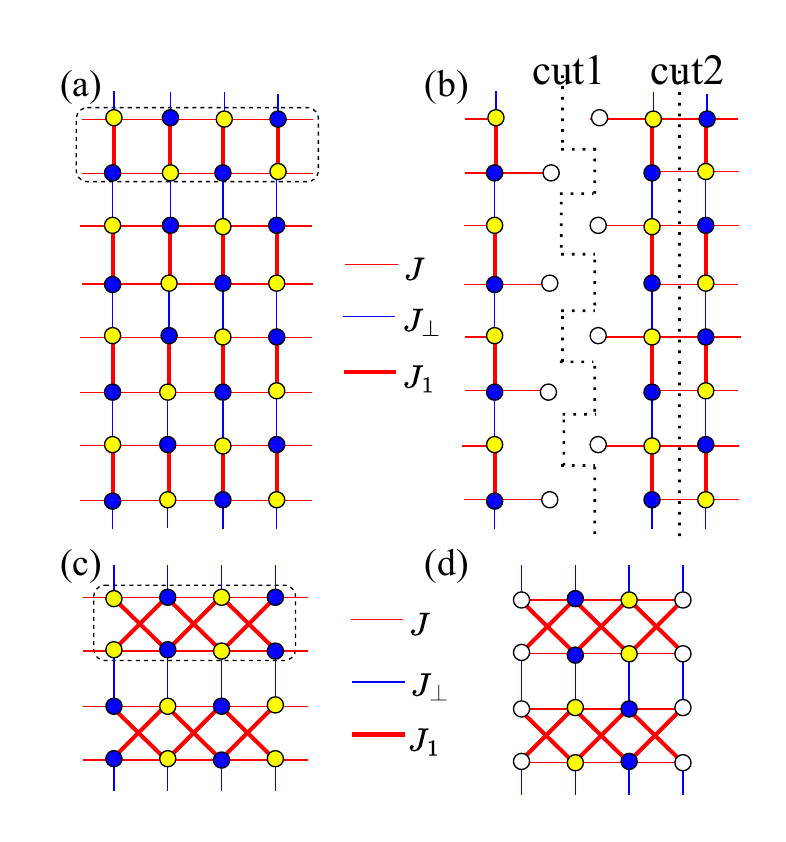}
	\caption{  (a) and (b) show the dimerized spin-$1/2$ Heisenberg model on a bipartite 
	square lattice with sublattices A (yellow circles) and B (blue circles).  Strong bonds $J_1>0$ are marked by thick red lines, weak bonds $J>0$ by 
	thin red lines, and interladder coupling $J_{\perp}>0$ by thin blue lines.
	(a) Periodic boundary conditions are applied in $x$ and $y$ directions. The dashed rectangular box encircles a usual ladder with AF rungs. 
	(b) Cutting along the dashed lines to expose edges, generating spinons along the edges opened by cut1, shown as open circles.
	Cut2 opens the boundaries trivially without spinons generated.  
	(c) and (d) show the Q1D coupled diagonal ladders, which are bipartite with sublattices A (yellow circles) and B (blue circles). 
	A diagonal ladder is shown inside the dashed rectangular box.
	(c) applies periodic boundary conditions in both $x$ and $y$ directions. 
	(d) applies open boundaries in $x$ direction to expose edges (open circles), with periodic boundary conditions $y$ direction.
	}
	\label{Fig:modelU}
    \end{figure}

The columnar dimerized spin-$1/2$ Heisenberg model on a square lattice, realized in Cu compounds such as BaCuSi$_2$O$_6$\cite{Okada2016Quasi}, undergoes a quantum phase transition from the dimer phase to the antiferromagnetic N\'eel phase (see Fig. \ref{Fig:modelU}(a)) and serves as a prototype for quantum criticality\cite{Sachdev2011}.
The dimer state has often been regarded as a typical example of a topologically trivial state, as evidenced by its gapped edge states \cite{Matsumoto2001, Ding2018}. 
However, if one views the model as a Q1D coupled usual ladder with AF rungs, then the gapped dimer state could be an SPT, as it is adiabatically connected to the 1D Haldane phase.
In this paper, we will solve this puzzle by demonstrating the emergence of a gapless ferromagnetically (FM) ordered edge state on a special zigzag edge, separated by cut1 in Fig. \ref{Fig:modelU}(b).
Furthermore, the strange correlator has been proposed as a direct probe of the SPT character without cutting the system in Q1D or 2D \cite{You2014, Wierschem2014, zhou2022detecting, Wierschem2014S}.
To furnish direct evidence of the SPT nature of the dimer state, we will generalize the strange correlator in accordance with its underlying theoretical principle. 
Utilizing them, we will provide additional evidence that the dimer state is an SPT.

It is then natural to expect the ordered edge state leading to extraordinary surface critical behaviors (SCBs) at the bulk critical point, which separates the dimer state from the AF N\'eel state and belongs to the (2+1)D O(3) universality class.  
This is a solely quantum mechanical effect, since, according to quantum-classical mapping, the (1+1)D surface can not host a long-range order for a system with continuous symmetry, and, therefore, an extraordinary SCB is excluded, but an extraordinary-log SCB is possible at the bulk critical point\cite{Metlitski2020}.
We then numerically study the surface critical behaviors at the bulk critical point. Our results prove the existence of an ordered surface state at the critical point, and the extraordinary SCBs are observed.

The paper is organized as follows. We describe the dimerized spin-1/2 Heisenberg model on a square lattice and the 
ground state phase diagram of the model in Sec. \ref{systems}.
In Sec. \ref{SPT}, we define generalized strange correlators to diagnose SPT in the even and odd topological sectors, respectively. 
The validity of these correlators is verified for the SPTs of the diagonal ladder and the usual ladder. 
Using these generalized strange correlators, we provide direct evidence that the dimer phase of the dimerized Heisenberg model is an SPT.
The nontrivial edge state of this SPT is also studied in this section. 
We investigate the SCBs in Sec. \ref{scbs}. We conclude in Sec.\ref{conclusion}.

\section{ Models and ground state phase diagram}  
\label{systems}
The dimerized Heisenberg model on a square lattice can be viewed as antiferromagnetically coupled spin-1/2 Heisenberg chains with alternating intrachain couplings. 
The Hamiltonian is written as
\begin{equation}
	\begin{split}
		H &= J_{1}\sum\limits_{i,j}\mathbf{S}_{i,2j}\cdot\mathbf{S}_{i,2j+1}+J_{\perp}\sum\limits_{i,j} \mathbf{S}_{i,2j+1}\cdot\mathbf{S}_{i,2j+2}\\
		&	+	J\sum\limits_{i,j}\mathbf{S}_{i,j}\cdot\mathbf{S}_{i+1,j},
	\end{split}
\end{equation}
where $\mathbf{S}_{i,j}$ is the spin at the $j$-th site in the $i$-th chain. $J_1>0$ and $J_{\perp}>0$ are intrachain couplings,  $J>0$ is the interchain coupling, see  Fig. \ref{Fig:modelU}(a).
We set the strong bond coupling $J_1=1$ to fix the energy scale. 

The system can also be viewed as a Q1D coupled usual ladders:
\begin{equation}
	\begin{split}
	H &= \sum_{j=0}H_{j}+ J_{\perp}\sum_{i,j=0}\mathbf{S}_{i,2j+1}\cdot \mathbf{S}_{i,2j+2},
		\end{split}
	\end{equation}
where the first sum is over the usual ladders with $H_{j}$ describing the $j$-th ladder,
	\begin{equation}
	\begin{split}
	H_j &= J \sum_{l=0,1} \sum_i \mathbf{S}_{i,2j+l}\cdot \mathbf{S}_{i+1,2j+l}+ J_1\sum_{i} \mathbf{S}_{i,2j}\cdot \mathbf{S}_{i,2j+1},
		\end{split}
		\label{H_ladder}
	\end{equation}
and the second sum describes the AF coupling between ladders. 

In Ref.\cite{Wang2022, Wangz2023},  we studied the CDLs, which is constructed by coupling the spin-1/2 diagonal ladders \cite{Kim2000}  either antiferromagnetically or ferromagnetically to form a 2D lattice, as illustrated in Fig. \ref{Fig:modelU}(c) for the AF case $J_\perp >0$. 
When interladder coupling is in the region of  $0 \leq J_\perp < (J_\perp)_c=0.17425(3)$ \cite{Wang2022} or $(J_\perp)_c=-0.18271(6)<J_\perp <0$ \cite{Wangz2023}, the system stays in the Haldane phases.  
At $(J_\perp)_c>0$, the system undergoes a phase transition to the AF phase (or to the striped phase at $(J_\perp)_c<0$), which belongs to the (2+1)D O(3) universality class.
We found gapless edge modes formed at the ladder endpoints (open circles in  Fig. \ref{Fig:modelU}(d) ) in the Haldane phases,  indicating  that the phases are
topologically nontrivial SPT \cite{Wang2022, Wangz2023}. 

\begin{figure}[h]
	\centering
	\includegraphics[width=0.46\textwidth]{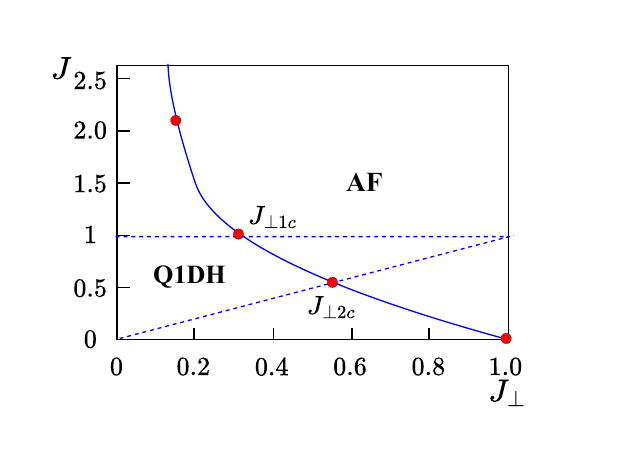}
	\caption{The $J$-$J_{\perp}$ phase diagram of the dimerized spin-1/2 Heisenberg model on a square lattice. The dimer phase is labeled by Q1DH, indicating that it is a Q1D Haldane state. 
	The dashed lines denote $J=1$ and $J=J_{\perp}$, respectively. 
$J_{\perp 1c}$ lables the critical point $(J=1, J_{\perp}=0.31407(5))$  and $J_{\perp 2c}$ labels the one $(J=0.52337(3),J_{\perp}=0.52337(3))$. }
	\label{Fig:diagramu}
\end{figure}

The ground state phase diagram of the dimerized Heisenberg model, or equivalently, the Q1D coupled usual ladders, has been 
obtained using the QMC simulations\cite{Matsumoto2001}, as shown in Fig.\ref{Fig:diagramu}. 
It is known that the ground state of a usual ladder is a Haldane state, which is a topologically nontrivial SPT state with string order \cite{White1996}.
Similar to the CDLs, the Haldane state is robust against weak higher-dimensional couplings between ladders due to the Haldane gap. 
The resulting dimer state is also referred to as a Q1D Haldane (Q1DH) state. 
The transition from the Q1DH state to the AF state belongs to the (2+1)D O(3) universality class.
There is no other phase transition when ($J$, $J_{\perp}$) sit on the left of the phase transition line between the AF state and the dimer state.
Thus, the Q1DH state is adiabatically connected to the SPT state of the usual ladder.  We, therefore,  expect the Q1DH state to also be a topologically 
nontrivial SPT phase. 
Nontrivial surface/edge states are expected, which may lead to different surface critical behavior compared to the classical phase transition in the 3D O(3) universality class.

The models studied in this work are all free of magnetic frustration and can be studied efficiently using quantum Monte Carlo (QMC) simulations. 
In this work, we employ the stochastic series expansion (SSE) QMC method with the loop algorithm \cite{Sandviksusc1991, Sandvik1999, Syljuasen} to study the 
edge properties of the model. To approach the ground-state thermodynamic limit as $L \to \infty$, the inverse temperature is set as $\beta=2L$, considering the dynamic critical exponent $z=1$ here.

\section{SPT and its edge modes} 
\label{SPT}
\subsection{Strange correlator: detect bulk SPT state without a cut} 
In 1D, the SPT states are characterized by string order \cite{den1989, Kim2000}.
For two-leg ladders, the ground states divide into two topologically distinct classes, described by two types of string order parameter: 
$\mathcal{S}_{\rm even}$ and $\mathcal{S}_{\rm odd}$, respectively. This is because the ground states of ladders are made up of short-range valence bonds (VBs) \cite{Kim2000}, and the two topological classes can be
distinguished by counting the number of VBs crossing an arbitrary line, which is always even for the usual ladder with AF rungs, but odd for the 
diagonal ladder, as illustrated in Fig. \ref{Fig:twoladders}(c) and (d). 
For the even sector, triplets tend to form on diagonals; therefore, 
the string order parameter  $\mathcal{S}_{\rm even}$ is defined using the 
spin-1 ${\mathbf S}^{(\rm E)}_i={\mathbf S}_{i,0}+{\mathbf S}_{i+1,1}$, Whilst for the odd sector, there is a tendency 
towards triplets on rungs, and the string order parameter  $\mathcal{S}_{\rm odd}$ is defined using the spin-1 ${\mathbf S}^{(\rm O)}_i=
{\mathbf S}_{i,0}+{\mathbf S}_{i,1}$, see  Fig. \ref{Fig:twoladders}(c) and (d).
Note that the spin-1s are formed by two nearest spin-1/2s on the same sublattices. 

However, the string order parameter is fragile to arbitrary weak higher-dimensional couplings between such chains or ladders \cite{Pollmann2010, Anfuso2007}, 
which has been recently verified numerically in Q1D coupled spin-1 Haldane chains model\cite{Zhu2021} and in Q1D spin-1/2 CDLs 
model \cite{Wang2022}.  To demonstrate the 2D/Q1D SPT state, the so-called strange correlator \cite{You2014,zhou2022detecting} is proposed
\begin{equation}
	\begin{split}
		C_{\rm SC}(r, r^\prime) = \frac{\langle
\Omega|\phi(r) \phi(r^\prime) | \Psi \rangle}{\langle \Omega|
\Psi\rangle},
		\end{split}
\end{equation}
in which $|\Psi\rangle$ is the ground state to be studied
and $|\Omega\rangle$ is a trivial product state.
Thus strange correlator describes a correlation function at the temporal domain wall between the states $|\Psi\rangle$ and $|\Omega\rangle$. 
You $et$ $ al.$ \cite{You2014} have shown that if $|\Psi\rangle$ is a nontrivial SPT state, then for some local
operator $\phi(r)$, $C_{\rm SC}(r, r^\prime)$ will either saturate to a constant in 1D or 2D, or at least decay as a power-law in 2D.

   \begin{figure}[h]
	\centering
	\includegraphics[width=0.46\textwidth]{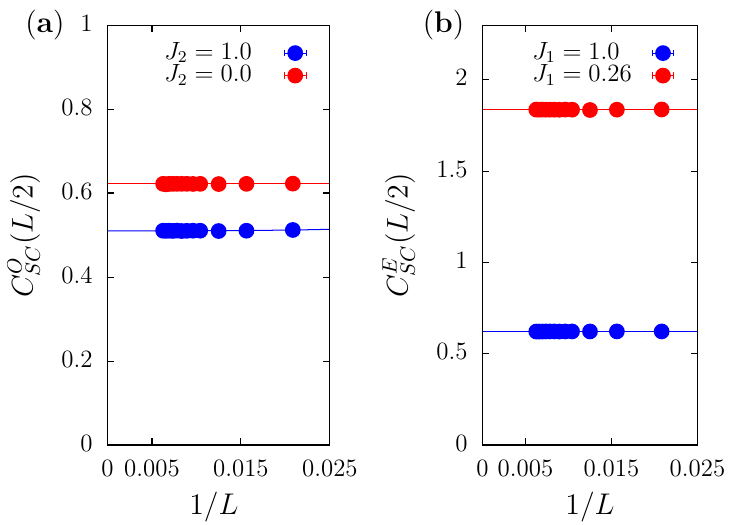}
	\caption{ Finite-size behavior of the strange correlators of the 1D diagonal and usual ladder for different parameters defined in Fig. \ref{Fig:twoladders}. 
	   (a) $C^{\rm O}_{\rm SC}(L/2)$ of the diagonal ladder with $J=J_{1}$=1, which converges to  0.5098(2) for $J_{2}=1.0$ and to  0.6217(4) for$J_{2}=0.0$, 
	   respectively. 
	   (b) $C^{\rm E}_{\rm SC}(L/2)$ of the usual ladder with $J$=1,  which converges to  0.62176(2) for $J_{1}=1.0$ and to 1.837(1) for $J_{1}=0.26$, respectively. }
	\label{Fig:1Dstrange}
\end{figure}

Here, we generalize the definition of the strange correlator in accordance with the underlying principle of the strange correlator.
The trivial state $|\Omega\rangle$ is chosen to be a VB state formed by a product of dimers in the topological sector, different from the 
ground state, which needs diagnosis. Meanwhile, the local operator is chosen to be the spin-1 operator of the system under investigation. 

For example, to demonstrate an SPT state in the odd sector, e.g., the ground state of the diagonal ladder, 
we can choose the local operator to be the spin-1 operator $\mathbf{S}^{(\rm O)}_i$ and
the product state $|\Omega^{(\rm E)}\rangle$ to be the VB state forming a columnar pattern, as shown in  Fig. \ref{Fig:twoladders} (f), which is topologically even.

To study the ground state of the usual ladder made up of VB configurations in the even sector, we chose the local operator to be 
$ \mathbf{S}_i^{(\rm E)}$ and the product state $|\Omega^{(\rm O)}\rangle$ to be the VB state forming a staggered pattern, as shown in  
Fig. \ref{Fig:twoladders} (e), which is topologically odd.
Therefore, two strange correlators $ C_{\rm SC}^{\rm O}(i, j)$ and $ C_{\rm SC}^{\rm E}(i, j)$ are defined for the SPT in the odd sector and the even sector, 
\begin{equation}
	\begin{split}
		C_{\rm SC}^{\rm O, E}(i, j) &= \frac{\langle \Omega^{(\rm E, O)} |\mathbf{S}^{(\rm O, E)}_{i} \cdot \mathbf{S}^{(\rm O, E)}_{j}| \Psi\rangle}
		{\langle\Omega^{(\rm E, O)}|\Psi\rangle},
		\end{split}
		\label{c_sc}
\end{equation} 
respectively.

Using the projector QMC method in the valence bond basis \cite{Sandvik2005, Sandvik2010}, we are able to directly access the
strange correlator defined in Eq.(\ref{c_sc}), through loop structure of the transposition graph. 
For example, $C_{\rm SC}^{\rm O}(i, j)$ can be estimated using the loop structure of the transposition graph Fig.\ref{Fig:twoladders}(g) where 
bonds in Fig. \ref{Fig:twoladders}(c), which is a typical VB state that constitutes the ground state $|\Psi\rangle$ of the usual ladder with AF rungs, and (e) 
which is the product state $|\Omega^{(\rm O)}\rangle$, are superimposed \cite{Sandvik2010}. 
Similarly, $C_{\rm SC}^{\rm E}(i, j)$ can be estimated using the loop structure of the transposition graph Fig.\ref{Fig:twoladders}(h) where 
bonds in Fig. \ref{Fig:twoladders}(d), which is a typical VB state that constitutes the ground state $|\Psi\rangle$ of the diagonal ladder, and (f) 
which is the product state $|\Omega^{(\rm E)}\rangle$ are superimposed.
Typically, as shown in Fig. \ref{Fig:twoladders}(g) and (h), for two spin-1s at the farthest distance, there is at least a pair of 
spin-1/2 spins in one loop that percolate the ladder, leading the correlator to be finite.  

In the simulations of a system of size $L$ with periodic boundaries, we calculate $C_{\rm SC}^{\rm O, E}(L/2)$ 
by averaging $C_{\rm SC}^{\rm O, E}(i, j)$  at the maximum available distance $|i-j|=L/2$ along an individual ladder, which can be viewed as the 
strange order parameter for the SPT in the odd and even sectors, respectively. 
(Strictly speaking, the strange correlator is not a genuine order parameter. However, for magnetically disordered states, it serves as an SPT indicator, as discussed in Appendix \ref{append:SC}).

For the 1D case, the coupling between ladders $J_{\perp}=0$. 
We calculate $C_{\rm SC}^{\rm O}(L/2)$ for the diagonal ladder and $C_{\rm SC}^{\rm E}(L/2)$ for the usual ladder. We observe that both of them are finite when $L$ goes to infinity,
as shown in Fig. \ref{Fig:1Dstrange} (a) and (b), for different rung or diagonal couplings.
 
We then calculate the strange order parameter in 2D. The Q1D coupled ladder systems are strongly anisotropic; therefore, we set the aspect ratio to $R=L/L_{\perp}=4$, where $L=L_x$ is the size along the ladder direction and $L_\perp=L_y$ is the size in the vertical direction. 
The linear size is reached up to $L=352$. As demonstrated in Appendix \ref{aspect_ratio}, the critical properties of the AF-dimer transition are not affected by the choice of $R$. 

\begin{figure}[h]
	\centering
	\includegraphics[width=0.46\textwidth]{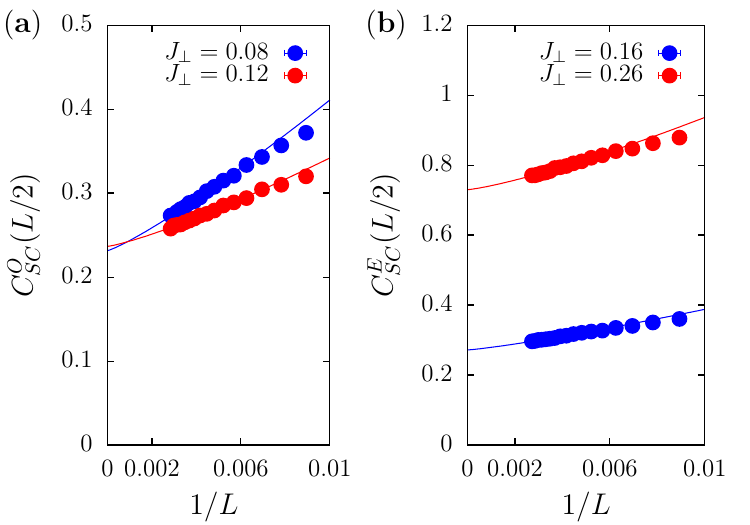}
	\caption{ Finite-size scaling analyses of the strange correlators of the Q1D coupled diagonal and usual ladders. 
	   (a) $C_{\rm SC}^{\rm O} (L/2)$ of the coupled diagonal ladders converges to 0.23(2) for $J_{\perp}=0.08$ and to 0.237(5) for $J_{\perp}=0.12$, 
	  with $J=J_1=J_2$=1. 
	   (b)  $C_{\rm SC}^{\rm E}(L/2)$  of Q1D coupled usual ladders  
	   converges to  0.27(2) for ($J=1$, $J_{\perp}=0.16$) and to 0.72(2) for  ($J=0.26$, $J_{\perp}=0.26$) with $J_1=1$. }
	\label{Fig:2Dstrange}
\end{figure}

The CDLs with $J=J_1=J_2=1$ and weak interladder couplings $J_{\perp} \in [0, 0.17425]$ is in the Q1D Haldane phase\cite{Wang2022, Wangz2023}.  
We find that the strange order parameter $C_{\rm sc}^{\rm O}(L/2)$ converges to a finite value when $L\to \infty$ in the Q1D Haldane phase, as 
shown in  Fig. \ref{Fig:2Dstrange} (a).
This provides direct evidence of a nontrivial SPT order in the Q1D Haldane phase of the weakly coupled diagonal ladders.
Similarly, as shown in  Fig. \ref{Fig:2Dstrange} (b), 
the strange order parameter $C_{\rm SC}^{\rm E}(L/2)$ also scales to a finite value when $L\to \infty$ in the dimer phase 
of the coupled usual ladders, which is also known as the dimersized spin-1/2 Heisenberg model. 
This result, combined with the fact that the dimer phase is adiabatically connected to the 1D usual ladder, suggests that the dimer phase of the dimerized sipn-1/2 Heisenberg model is an SPT, which is topologically nontrivial.

\subsection{surface state}

\begin{figure}[htb]
	\centering
	\includegraphics[width=0.38\textwidth]{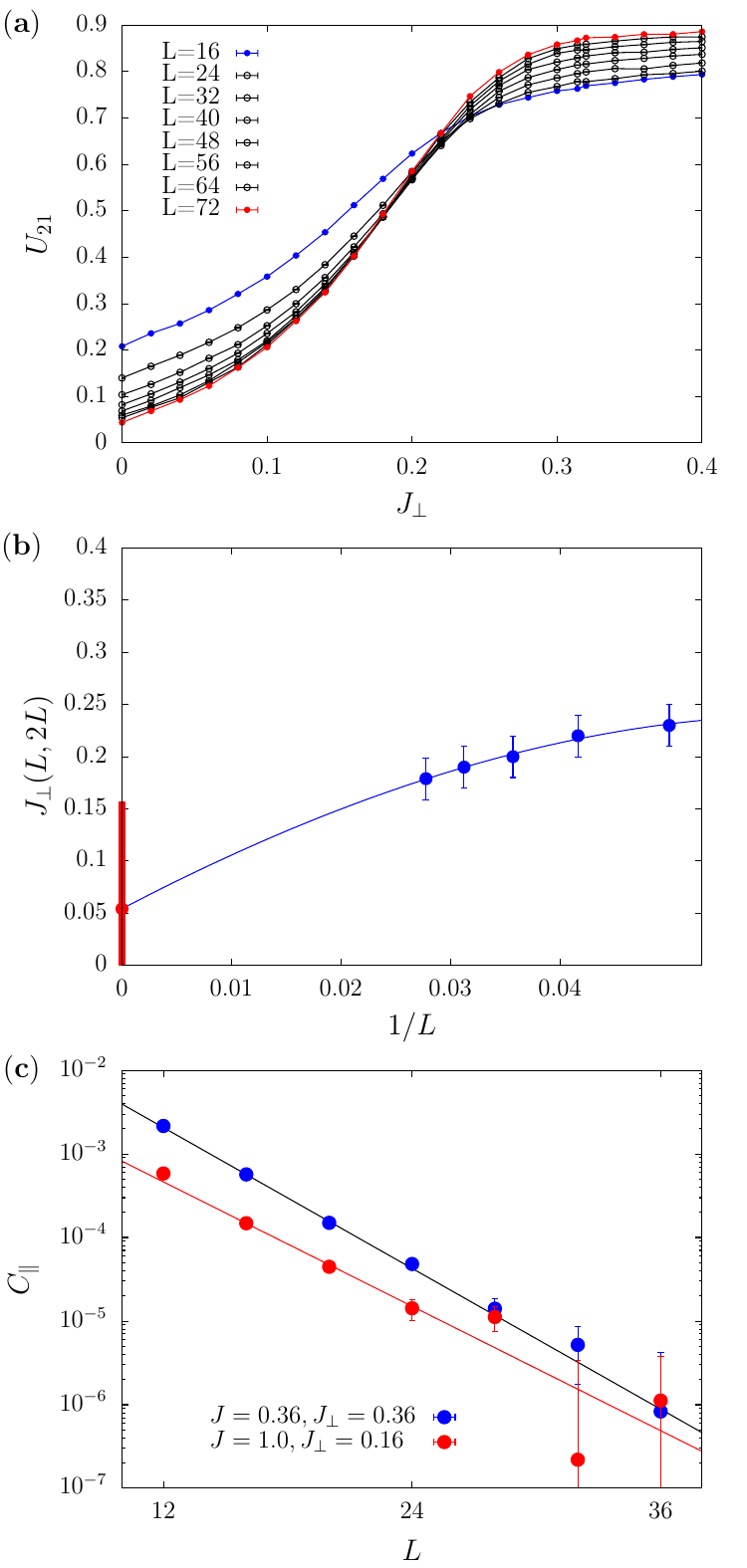}
	\caption{Properties of the zigzag edge and the vertical edge in the gapped bulk Q1DH phase of the dimerized Heisenberg model.  
	(a) The surface Binder cumulant $ U_{21} $ of the zigzag edge as a function of $ J_{\perp} $ for various system sizes with $ J=J_1=1 $. 
	(b) The crossings $J_{\perp}(L,2L)$ of $U_{21}(L)$ and $U_{21}(2L)$ are plotted as a function of the inverse of system size  $1/L$.  
	The red point represents the extrapolated crossing point for $L\to \infty$, indicating that, 
	as the system size increases, the crossings drift toward $ J_{\perp} =0$ and the edge is FM ordered as far as $J_\perp>0$.
	(c)  $C_{\parallel}(L/2)$ on the vertical edge with ($J=1$, $J_{\perp}=0.16$) and ($J=0.36$, $J_{\perp}=0.36$) versus size $L$ on a 
	linear-log scale. Exponential decay with  $L$ is observed, meaning that the edge states are gapped. }
	\label{Fig:bgapsbinder}
    \end{figure}

  The hallmark of SPT states is the presence of nontrivial surface states that are either gapless or degenerate in the gapped bulk state\cite{wen2017}. 
  In 1D, this can be understood through the short-range VB picture of the SPT state. As shown in Fig. \ref{Fig:twoladders}(d) for the diagonal ladder, when the periodic ladder is cut to open two ends, the VB connecting two neighboring $S=1$ spins is broken and leaves two spinons at the ends, leading
  to four degenerate edge states.  
  For Q1D coupled diagonal ladders,  cutting the ladders perpendicular to the ladder direction exposes two boundaries with the VBs connecting neighboring spin-1s
  cutted but the spin-1s stay untouched. The spinons generated at the ends of the ladders form two dangling spin-1/2 chains at the boundaries, 
  as illustrated in Fig. \ref{Fig:modelU}(d), which is gapless.
  This picture, as expected for an SPT, has been demonstrated numerically in \cite{Wang2022, Wangz2023}.

  For the usual ladder with AF rungs, cutting the ladder vertically to open two ends breaks an even number of VBs, see `cut2' in Fig. \ref{Fig:twoladders}(c), we see no spinon is created at the ends.
  However, it is possible to have two spinons at the ends of a cut ladder, if one cuts the ladder diagonally, as illustrated by `cut1' in Fig. \ref{Fig:twoladders}(c), such that one VB connecting two neighboring $S=1$ spins is cut and leaves one spinon on each end. Note that such cutting 
  does not touch the spin-1s formed diagonally. 
  It is thus natural to expect that the edges exposed by vertically cutting the ladders (cut2) are gapped. In contrast, the edges exposed by cutting the coupled usual ladders in a zigzag way (cut1), shown 
  in Fig. \ref{Fig:modelU}(b), consist spinons thus generated, forming dangling spin-1/2 chains. 
  One important fact is that the spinons always reside on one sublattice on one side of the cut, but on the other sublattice on the other side. 
 This leads to effective ferromagnetic interactions between neighboring spinons, resulting in long-range order at the edges.
 
We now numerically verify the above expectations.  
To nail down the long-range order on the zigzag boundary shown in Fig. \ref{Fig:modelU}(b), we calculate the surface Binder cumulant $U_{21}$ \cite{Binder1981, Binder1984},
\be
 U_{21}(L)= \frac{5}{6}\left(3-\frac{\langle m_{1}(L)^{4}\rangle}{\langle m_{1}(L)^{2}\rangle^{2}}\right).
 \ee
where $m_{1}(L)$ is the boundary magnetization defined as $m_{1}(L) =\frac{1}{L}\sum_{i \in {\rm boundary}}  S_i^z$,
with the summation restricted for spins on the boundary. 
$U_{21}(L)$ converging to 1 with $L \to \infty$ indicates the existence of the magnetic order, and approaching
zero with increasing system size implies that the system is in the magnetically disordered state.

As shown in Fig. \ref{Fig:bgapsbinder}(a), it seems that there is a surface transition from disorder to ordered states at a finite $J_\perp$. 
However, the crossings of $U_{21}$ for a pair of sizes $L$ and $2L$ drift towards small values of $J_\perp$. 
Fitting a polynomial $J_{\perp}(L,2L)=c +a_{1}L^{-1}+a_{2}L^{-2}+a_{3}L^{-3}$ to the crossings  
 yields $c=0.05 \pm 0.1$, which is zero within one error bar, 
 as shown in Fig. \ref{Fig:bgapsbinder}(b). 
This indicates that, as far as $J_\perp$ is nonzero, the ferromagnetic order appears on the boundary.

On the contrary, we show that the boundary indicated by cut2, see Fig. \ref{Fig:modelU}(b), is gapped by calculating the  
boundary parallel correlation $C_\parallel(L/2)$ between two edge spins $S_i^z$ and $S_j^z$ at the longest distance $|i-j|=L/2$.  
We observe an exponential decay of $C_\parallel(L/2)$ at the edge cut 2.  As 
shown in Fig. \ref{Fig:bgapsbinder} (c), the data in the dimer phase can be fitted using straight lines on a linear-log scale, meaning that the correlation decays exponentially. 
Fitting the curves with
\begin{equation}
    C_\parallel(L/2) \sim \exp{(-L/a)},
    \label{surface_gap}
\end{equation}
we obtain $a=3.09(8)$ at ($J=J_{\perp}=0.36$) and $a= 3.5(4)$ at ($J=1$, $J_{\perp}=0.16$).
The edge states on the cut-2 are indeed gapped.

\section{ Surface critical behavior}  
\label{scbs}
We now investigate the surface critical behavior at the bulk critical point of the model. 

Since the zigzag edge (or surface) is ordered for any $J_\perp>0$ due to the bulk-edge correspondence of SPT, which is a quantum mechanical effect, we naturally expect the associated SCB to belong to the 
extraordinary transition, which is characterized by anomalous surface scaling dimension $\eta_\parallel$ and the surface magnetic field scaling dimension $y_{h1}$
of the surface field $h_1$.
In particular, at bulk critical points, the finite-size scaling behaviors of the $C_\parallel(L/2)$ and $m_{1}^{2}(L)$ 
behave as 
\begin{equation}
C_{\parallel}(L/2) \sim C_{\parallel}+ a_1 L^{-(d+z-2+\eta_{\parallel})},
\label{cs1}
\end{equation}
and
\begin{equation}
m_{1}^{2}(L) \sim m_{1}^{2} + a_2 L^{-(2d+2z-2-2y_{h1})},
\label{m1s}
\end{equation}
respectively, 
where $C_\parallel=m_{1}^{2}\neq 0$ characterizes a long-range order on the surface,
$\eta_\parallel$ and  $y_{h1}$ describe the extraordinary SCBs, $a_1$ and $a_2$ are unknown constants  \cite{Binder1974}.

\begin{figure}[htb]
	\centering
	\includegraphics[width=0.5\textwidth]{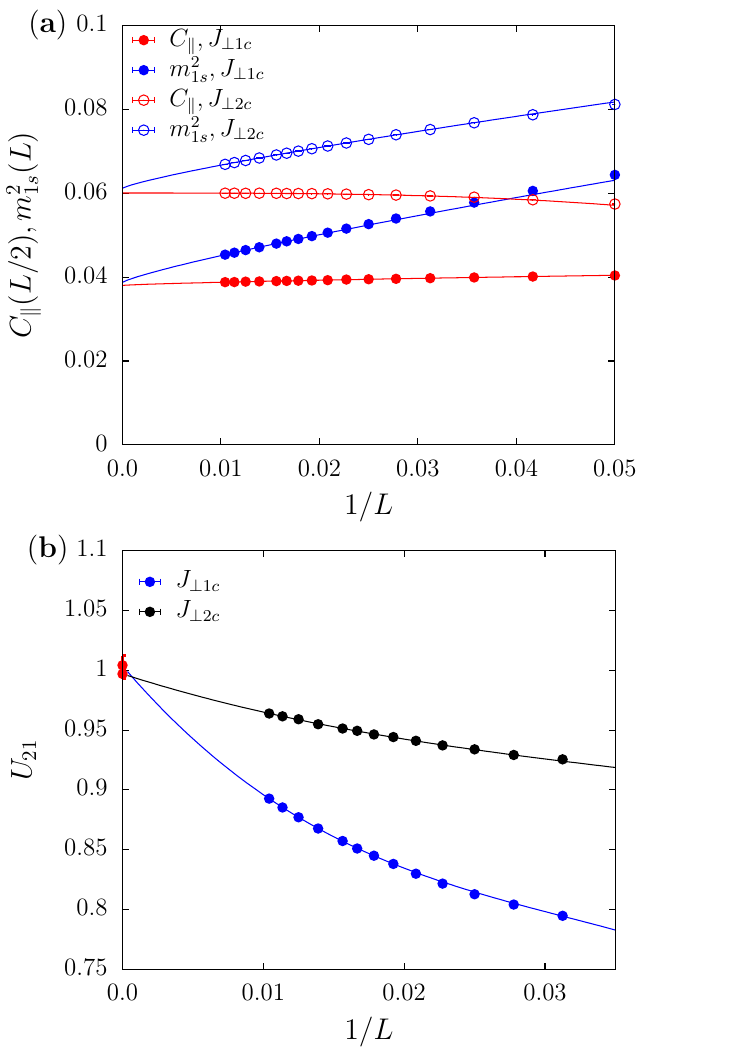}
	\caption{(a) The correlations $C_{\parallel}(L/2)$   and squared surface magnetization $m_1^2$ versus $1/L$  at  bulk critical points $J_{\perp 1c}$ and $J_{\perp 2c}$. 
	 (b) Surface Binder cumulant $U_{21}$ for different system sizes at bulk critical points $J_{\perp 1c}$ and $J_{\perp 2c}$. The red symbol shows the extrapolated value in the thermodynamic limit, which is 1 within the error bar.}
	\label{Fig:scb}
    \end{figure}

The numerical results of  $C_\parallel(L/2)$ and $m^2_{1}(L)$ versus $1/L$ at two representative bulk critical points $J_{\perp 1c}$ and $J_{\perp 2c}$ 
are drawn in Fig. \ref{Fig:scb}(a).

Fitting these results according to Eqs. (\ref{cs1}) and (\ref{m1s}) at $J_{\perp 2c}$, 
we find $C_\parallel=0.0599(3)$, $m^2_{1}=0.061(1)$ which are consistent with each other. This indicates the existence of a long-range order on the surface. However, it is difficult to obtain a meaningful estimate of $\eta_\parallel$; we only obtain  $y_{h1}=1.57(6)$.

We then study the SCBs at $J_{\perp 1c}$. 
Since the critical points $J_{\perp 1c}$ and $J_{\perp 2c}$ are sitting on the same phase transition line, belonging to the same universality class: (2+1)D O(3). 
On the same zigzag edge, we expect qualitatively the same SCBs at $J_{\perp 1c}$ as those at the $J_{\perp 2c}$. 
Fitting the results at $J_{\perp 1c }$ according to Eqs. (\ref{cs1}) and (\ref{m1s}), we find $C_\parallel=0.0380(3)$, $m^2_{1}=0.039(2)$, 
which are consistent within the error bars, 
thus also supporting a long-range order on the surface and suggesting extraordinary SCBs. 
In addition to $y_{h1}=1.58(7)$, we find  $\eta_{\parallel}=-0.27(6)$ at $J_{\perp 1c}$. These two exponents satisfy the scaling relation $\eta_\parallel = d+z - 2y_{h1}$ \cite{Diehl} and $y_{h1}$ is consistent with that obtained at $J_{\perp 2c}$.

To further nail down the long-range order on the zigzag edge at bulk critical points, we calculate the surface Binder cumulant $U_{21}$ at $J_{\perp 1c}$
and $J_{\perp 2c}$.
Numerical results of $ U_{21}(L)$ as a function of size $1/L$ are plotted in  Fig. \ref{Fig:scb} (b). We  fit the data via a polynomial of $1/L$
\begin{equation}
U_{21}(L) =U_{21}+c_{1}L^{-1}+c_{2}L^{-2}+c_{3}L^{-3},
\label{cs1p}
\end{equation}
and find statistically sound estimation $U_{21}=1.004(8)$ at $J_{\perp 1c}$ and $U_{21}=0.996(5)$ at $J_{\perp 2c}$ which are within one estimated error bar. This further supports the existence of a long-range order on the zigzag edge/surface.

\section{Conclusions}
\label{conclusion}
Using quantum Monte Carlo simulations, we have studied the dimerized spin-1/2 Heisenberg model on a square lattice, which is also viewed as two-dimensional antiferromagnetically coupled usual ladders with AF rungs.  
Inspired by the fact that the usual ladder with AF rungs and the diagonal ladder both exhibit topologically nontrivial Haldane states, but in different topological sectors, and the Q1D Haldane state of the coupled diagonal ladders is an SPT with a gapless edge state, we argued that the gapped dimer phase of the dimerized Heisenberg model, which was considered topologically trivial previously, should be a topologically non-trivial SPT with nontrivial edge states. 
Since the commonly used string order parameter describing 1D SPT states is fragile to arbitrary weak higher-dimensional couplings between such chains or ladders, we have defined odd and even generalized strange correlators to demonstrate an SPT state, which converge to a finite value for the SPT state in the odd and even topological sectors, respectively. 
Using the projector QMC simulations, the validity of the generalized strange correlator method has been verified in the diagonal ladder, usual 
ladder, and Q1D CDLs.  In this way, the dimer phase of the dimerized Heisenberg model is shown to be a topologically nontrivial SPT through the study of the newly introduced strange correlator.

To further prove that the dimer state is topologically nontrivial, we have studied the edge state on the edges exposed by a zigzag cut in the dimer state.
Using SSE QMC simulations, we have demonstrated that the edge state is FM ordered. The mechanism of this ordering is that spinons are generated on both sides of the zigzag cut, which is typical for an SPT state; these spinons stay on one sublattice on one side of the cut, but on the other sublattice on the other side of the cut, and therefore, are coupled by effective ferromagnetic interactions. 
These results provided solid evidence supporting that the dimer phase is a topologically nontrivial SPT state.

Since the zigzag edge is ordered for any $J_\perp>0$ due to quantum mechanical effects, we expect the edge to exhibit extraordinary surface critical behavior when the bulk is tuned to the critical point. We have studied the surface critical behavior on the zigzag edges of the dimerized spin-$1/2$ Heisenberg model and have confirmed the existence of the long-range order on these edges at the bulk critical point through large-scale QMC simulations. The surface, therefore, exhibits extraordinary surface critical behavior at the (2+1)-dimensional O(3) bulk critical point, in contradiction to theoretical predictions based on the classical-quantum mapping.

\begin{acknowledgments}
We acknowledge the helpful discussion with  Chao-Ming Jian.
Z.W., L.L., and W.G. thank the support from the National Natural Science Foundation of China under Grant Nos. 12175015 and 12574252. Z.L. and Z.Y. are supported by 
the Scientific Research Project (No.WU2024B027) and the Start-up Funding of Westlake University.
The authors also acknowledge the Super Computing Center of Beijing Normal University and Beijing PARATERA Tech Co., Ltd., and the HPC centre of Westlake University 
for providing HPC resources.  Z.W. is supported by the China Postdoctoral Science Foundation under Grant No.2024M752898. Z.L. thanks the China Postdoctoral 
Science Foundation under Grants No.2024M762935 and NSFC Special Fund for Theoretical Physics under Grants No.12447119.  S.Q.N. is supported by a CRF from the Research Grants
Council of the Hong Kong (No. C7037-22GF).
Y.C.W. acknowledges the support from the Natural Science Foundation of China (Grant No. 12474216) and Zhejiang Provincial Natural Science Foundation of China (Grant No. LZ23A040003), and the Start-up Funding of Hangzhou International Innovation Institute of Beihang University. 
\end{acknowledgments}

{\it{\color{blue} Data availability.-}}
The data that support the findings of this article are openly available\cite{wang2026zenodo}.

\appendix
\section{ Strange correlator serves as an SPT indicator only in magnetically disordered phase  }
\label{append:SC}

For a nontrivial SPT state, the strange correlator either saturates to a constant in 1D or 2D, or at least decays as a power law in 2D\cite{You2014}. However,  this behavior constitutes a necessary but not sufficient condition (an order parameter) for identifying an SPT state.  

\begin{figure}[htb]
\centering
\includegraphics[width=0.49\textwidth]{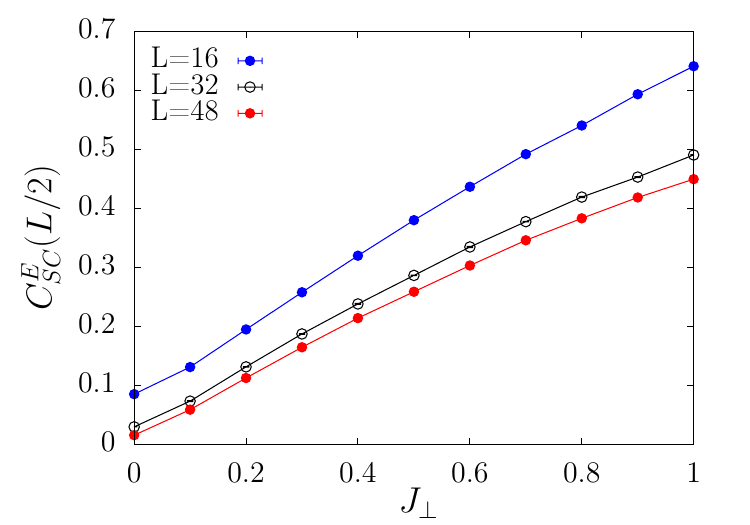}
\caption{ 
The strange correlator $C_{\rm SC}^{\rm E}(L/2)$ plotted as a function of $ J_1(=J_{\perp})$ for different system sizes in the Q1D coupled usual ladders.  The aspect ratio $ R = 4 $, and $J=1$ is fixed.
}
\label{fig:smstrange}
\end{figure}

To illustrate this, we present the strange correlator $C_{\rm SC}^{\rm E}(L/2)$ for the Q1D coupled usual ladders as a function of $ J_1(=J_{\perp})$ across different system sizes in Fig. \ref{fig:smstrange}.
At finite $J_\perp=J_1$, the system is magnetically ordered. We find that within this phase, the strange correlator also converges to a finite value in the thermodynamic limit.

However, as $J_\perp=J_1\to 0$, the system decouples into topological trivial $S=1/2$ Heisenberg chains,  $C_{\rm SC}^{\rm E}(L/2)$ converges to zero.  
Therefore, for a magnetically disordered phase, the strange correlator can distinguish a quasi-one-dimensional SPT phase and a trivial phase.

\section{Aspect ratio and critical properties of the dimer-AF transition}
\label{aspect_ratio}

The dimerized Heisenberg model is highly anisotropic, where correlation lengths along the ladder direction and the vertical direction may depend on the aspect ratio of the system. We present here additional simulation results of horizontal ($\xi=\xi_x$) and vertical ($\xi_{\perp}=\xi_y$) correlation lengths in systems with aspect ratios $ R=L/L_{\perp}=1$ and $ R=4$, respectively, to show that critical properties of the dimer-AF transition are not affected by the choice of aspect ratio.  In the simulations, the inverse temperature is set to $\beta= L$, with periodic boundary conditions
applied along both $x$ and $y$ lattice directions. 

The  correlation length is computed from the static spin structure factor (second-moment method \cite{SandvikAIP} ), 
\begin{equation}
S(\mathbf{q}) = \sum_{\mathbf{r}} e^{-i \mathbf{q} \cdot \mathbf{r}} C(\mathbf{r}),
\end{equation}  
where \( C(\mathbf{r}) = \langle S^z_{\mathbf{r}} S^z_{\mathbf{0}} \rangle \) is the spin-spin correlation function. The correlation length $\xi=\xi_x$ is given by  
\begin{equation}
\xi = \frac{L}{2\pi} \sqrt{ \frac{S(\pi, \pi)}{S(\pi + 2\pi/L, \pi)} - 1 },
\end{equation}  
and $\xi_{\perp}=\xi_y$ is given by
\begin{equation}
\xi_{\perp} = \frac{L_{\perp}}{2\pi} \sqrt{ \frac{S(\pi, \pi)}{S(\pi , \pi+ 2\pi/L_{\perp})} - 1 }.
\end{equation}

\begin{figure}[htb]
\centering
\includegraphics[width=0.49\textwidth]{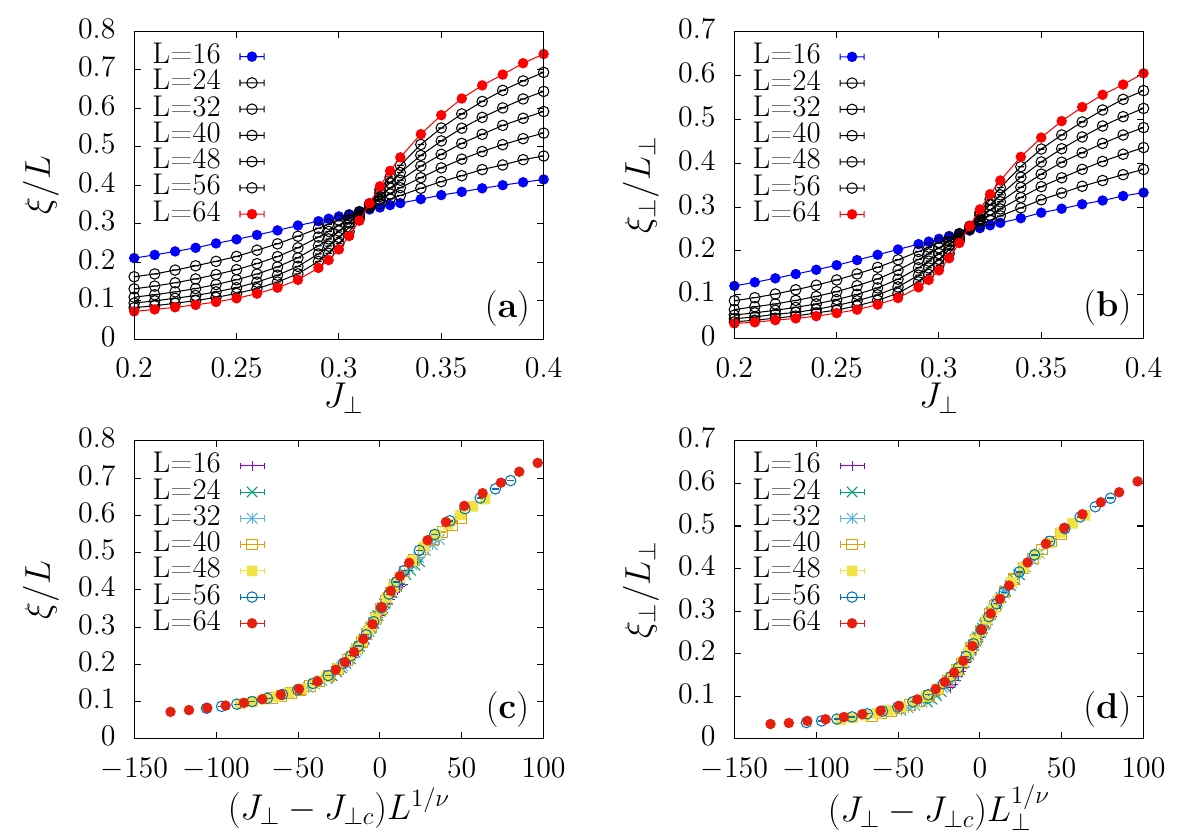}
\caption{  Results for $ R=1$. (a) $\xi/L$ and (b) $\xi_{\perp}/L_{\perp}$ versus $J_{\perp}$ for different system sizes.  Data collapse of $\xi/L$ and $\xi_{\perp}/L_{\perp}$ is shown in (c) and (d), respectively, using the critical exponent $\nu=0.71$ and $J_{\perp c}=0.31407$ for fixed $J=1$. Error bars are much smaller than the symbols. }
\label{fig:xir1}
\end{figure}

Any singular physical quantity $ A $ (not necessarily divergent) in the vicinity of a QCP obeys the finite-size scaling ansatz
\begin{equation}
	A(J_{\perp}, L) = L^{\kappa} f\left[(J_{\perp} - J_{\perp c}) L^{1/\nu}, L^{-\omega}\right],
\label{scaling}
\end{equation}  
where $ \nu $ is the correlation length exponent, $ f(x) $ is a scaling function, and $ \omega > 0 $ is the effective correction-to-scaling exponent.  
For $ A$ being the scaled correlation length $\xi / L $, or, $\xi_{\perp} / L_{\perp}$, which is dimensionless, we have $ \kappa = 0 $.

\begin{figure}[htb]
\centering
\includegraphics[width=0.49\textwidth]{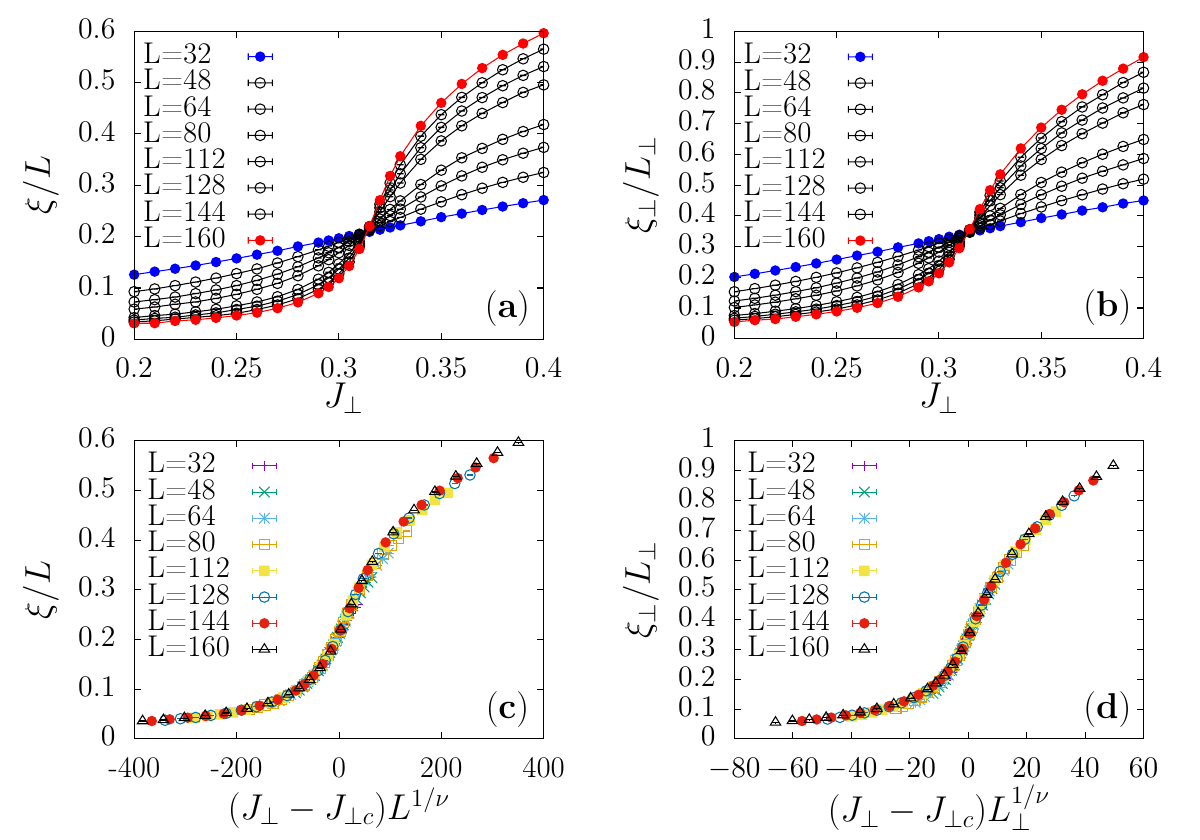}
\caption{ Results for $ R=4$. (a) $\xi/L$ and (b) $\xi_{\perp}/L_{\perp}$ versus $J_{\perp}$ for different system sizes $L=L_{x}$.  Data collapse of $\xi/L$ and $\xi_{\perp}/L_{\perp}$ is shown in (c) and (d), respectively, using $\nu=0.71$ and $J_{\perp c}=0.31407$ for fixed $J=1$. Error bars are much smaller than the symbols.}
\label{fig:xir4}
\end{figure}

Figure \ref{fig:xir1} (Fig. \ref{fig:xir4}) shows scaled correlation lengths $\xi/L$ and $\xi_\perp/L_\perp$ of different system sizes as a function of $J_{\perp}$  for $ R=1$ and $ R=4$, respectively. 
It is clear that, at the critical point--identified by the crossing of different system sizes--the horizontal and vertical correlation lengths $\xi$ and $\xi_\perp$ differ in magnitude for a given aspect ratio $R$. Furthermore, the correlation length $\xi$ and $\xi_\perp$ vary with $ R $. Nevertheless, the scaled correlation lengths under different settings all collapse onto the scaling function using the universal critical exponent of the (2+1)-dimensional O(3) universality class $\nu=0.71$ \cite{Matsumoto2001} and the critical point $J_{\perp c}=0.31407$, indicating that the choice of aspect ratio does not affect the critical properties of the model.

\bibliography{ref.bib}

\end{document}